\documentclass[twocolumn,showpacs,preprintnumbers,amsmath,amssymb]{revtex4}

\usepackage{graphicx}
\usepackage{dcolumn}
\usepackage{bm}

\begin{document} 

\preprint{}

\title{Epitaxial ferromagnetic Fe$_{3}$Si/Si(111) structures with high-quality hetero-interfaces}

\author{K. Hamaya,\footnote{hamaya@ed.kyushu-u.ac.jp} K. Ueda, Y. Kishi, Y. Ando, T. Sadoh, and M. Miyao\footnote{miyao@ed.kyushu-u.ac.jp}}
\affiliation{%
Department of Electronics, Kyushu University, 744 Motooka, Fukuoka 819-0395, Japan}%


%

\date{\today}
\begin{abstract}
To develop silicon-based spintronic devices, we have explored high-quality ferromagnetic Fe$_{3}$Si/silicon (Si) structures. Using low-temperature molecular beam epitaxy at 130 $^{\circ}$C, we realize epitaxial growth of ferromagnetic Fe$_{3}$Si layers on Si (111) with keeping an abrupt interface, and the grown Fe$_{3}$Si layer has the ordered $DO_\text{3}$ phase. Measurements of magnetic and electrical properties for the Fe$_{3}$Si/Si(111) yield a magnetic moment of $\sim$ 3.16 $\mu_\text{B}$/f.u. at room temperature and a rectifying Schottky-diode behavior with the ideality factor of $\sim$ 1.08, respectively. 
\end{abstract}

\maketitle
Semiconductor spintronic devices such as spin-field effect transistors (spin FET) are one of the possible candidates to substitute for existing silicon-based complementary metal-oxide-semiconductor devices.\cite{Wolf,Datta,Sugahara,Hall} To realize operations of the spin FET, an electrical spin injection from ferromagnets into semiconductors is an essential technology. For III-V semiconductor devices, several groups have demonstrated highly efficient spin injection and detection using an epitaxial Fe thin film and tailored Schottky tunnel barriers so far.\cite{Hanbicki,Hanbicki2,Lou} From these facts, it is necessary for semiconductor spintronics to develop crystal growth techniques of ferromagnets on semiconductors with keeping high-quality interfaces. In particular, it will become key to build epitaxial growth of ferromagnets on silicon (Si) from the viewpoint of application to existing silicon large-scale integrated circuit (LSI) technologies.\cite{Min} 

Moreover, for spintronics, Si has been regarded as an ideal material because of a long spin relaxation time due to weak spin-orbit interaction, weak hyperfine interaction and lattice inversion symmetry, which will give rise to a long spin diffusion length in the devices. Recently, spin transport in Si conduction channels was experimentally demonstrated although their operations were limited at low temperatures.\cite{Appelbaum,Jonker1,Jonker2} This means that the spin degree of freedom can be introduced into Si-based electronic devices. 

To date, ferromagnetic MnAs thin films have been grown epitaxially on Si (001),\cite{Tanaka} but electrical spin injection from MnAs into Si across a Schottky tunnel barrier has never been demonstrated unfortunately. Also, the Curie temperature of MnAs is $\sim$ 315 K,\cite{Plake} which may be relatively low for an operation temperature of future LSIs. Thus, possibilities of other high-Curie temperature materials compatible with Si should be explored. Here we select a ferromagnetic Heusler alloy Fe$_{3}$Si thin film, which has a high Curie temperature above 800 K, a relatively high spin polarization of $\sim$ 45 \% and a small coercive field of $\sim$ 7.5 Oe.\cite{Ionescu} In this letter, we achieve highly epitaxial growth of the Fe$_{3}$Si layer on Si (111) using low-temperature molecular beam epitaxy (MBE) at 130 $^{\circ}$C. The interface between Fe$_{3}$Si and Si keeps an abrupt interfacial structure and the grown Fe$_{3}$Si thin film has the ordered $DO_\text{3}$ phase. 
\begin{figure}[b]
\includegraphics[width=8.5cm]{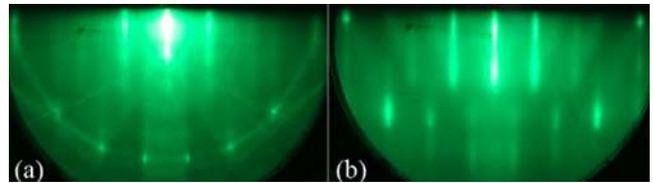}
\caption{(Color online) Reflection high energy electron diffraction (RHEED) patterns observed along the [$\bar{2}$11] azimuth for (a) Si (111) and (b) Fe$_{3}$Si surfaces after the surface cleaning at 450 $^{\circ}$C and the growth at 130 $^{\circ}$C, respectively.}
\end{figure}  

An n-type Si wafer with (111) orientation was used as the substrate to grow the ferromagnetic Fe$_{3}$Si. After cleaning the substrates with an aqueous HF solution (HF : H$_{2}$O $=$ 1 : 40), we conducted a heat treatment at 450 $^{\circ}$C for 20 min in an MBE chamber with a base pressure of 2 $\times$ 10$^{-9}$ Torr. A reflection high energy electron diffraction (RHEED) pattern of the flashed surface of the substrate displays symmetrical streaks, as shown in Fig. 1(a). This figure indicates atomic level flatness of the surface of the Si substrate. Prior to the growth of Fe$_{3}$Si layers, the substrate temperature was reduced down to 130 $^{\circ}$C. Using Knudsen cells, we co-evaporated Fe and Si with the growth rates of 2.12 and 1.20 nm/min, respectively. Figure 1(b) shows an RHEED pattern of the Fe$_{3}$Si layer observed. The RHEED pattern clearly keeps symmetrical streaks, which implies good two-dimentional growth of the Fe$_{3}$Si on Si(111). X-ray diffraction (XRD) measurements of the Fe$_{3}$Si/Si(111) were also performed in $\theta-2\theta$ and 2$\theta$ configurations. In general, since the lattice mismatch between Fe$_{3}$Si and Si is about 4 \% for bulk samples, the diffraction peaks can be separated. For our samples, however, the intensity of the diffraction peak due to Fe$_{3}$Si(111) was quite weaker than that due to Si(111), and then the Fe$_{3}$Si(111) peak was overlapped by the foot of the strong Si(111) peak. As a result, we could separate hardly the diffraction peak due to the Fe$_{3}$Si layer from that due to the Si substrate in $\theta-2\theta$ measurements.\cite{ref1} Also, there was no peak due to other Fe$_{x}$Si$_{y}$ compounds. From these results, we can judge that the Fe$_{3}$Si layer is grown epitaxially on the Si(111) substrate with no secondary phase such as cubic-FeSi.

\begin{figure}[t]
\includegraphics[width=8.5cm]{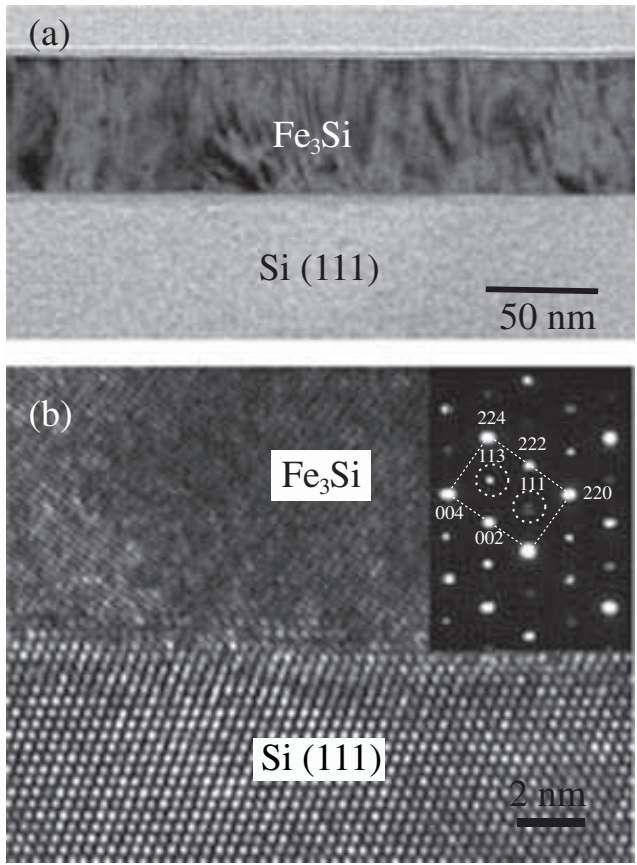}
\caption{Cross-sectional (a) low-magnification and (b) high-resolution TEM images of the Fe$_{3}$Si/Si(111) hybrid structure. The inset shows nanobeam electron diffraction patterns of the epitaxial Fe$_{3}$Si layer. The zone axis is parallel to the [1$\bar{1}$0] direction.}
\end{figure}  

Figure 2 (a) shows a cross-sectional transmission electron microscopy (TEM) image of the Fe$_{3}$Si/Si(111) structure. The interface between Fe$_{3}$Si and Si(111) yields almost no marked roughness and no reaction layer. Also, selected area diffraction measurements revealed the absence of interfacial reaction phases such as cubic-FeSi (not shown here). A high-resolution TEM image near the Fe$_{3}$Si/Si(111) interface is also shown in Fig. 2(b). The interface has a high-quality abruptness within the fluctuation of 2 $\sim$ 3 monolayers. We also performed Rutherford backscattering spectroscopy (RBS) measurements of the Fe$_{3}$Si/Si(111) and obtained the minimum yield ($\chi_\mathrm{min}$) of $\sim$ 12 \%. These results indicate that the Fe$_{3}$Si/Si(111) maintains high crystallinity comparable to the Fe$_{3}$Si/Ge(111) in our previous works.\cite{Sadoh,Maeda} Nanobeam electron diffraction patterns of the Fe$_{3}$Si layers are also presented in the inset of Fig. 2(b). We can see evident superlattice diffractions, (111) and (113), caused by the presence of the ordered $DO_\text{3}$ phase (white broken circles), in addition to fundamental reflections of the $A2 + B2 + DO_\text{3}$ phases and superlattice reflections of the ordered $B2 + DO_\text{3}$ phases. From these structural characterizations, the high-quality formation of the ordered single-crystal Fe$_{3}$Si layers has been realized on the Si substrate at a quite low temperature of 130 $^{\circ}$C.
\begin{figure}[t]
\includegraphics[width=8.5cm]{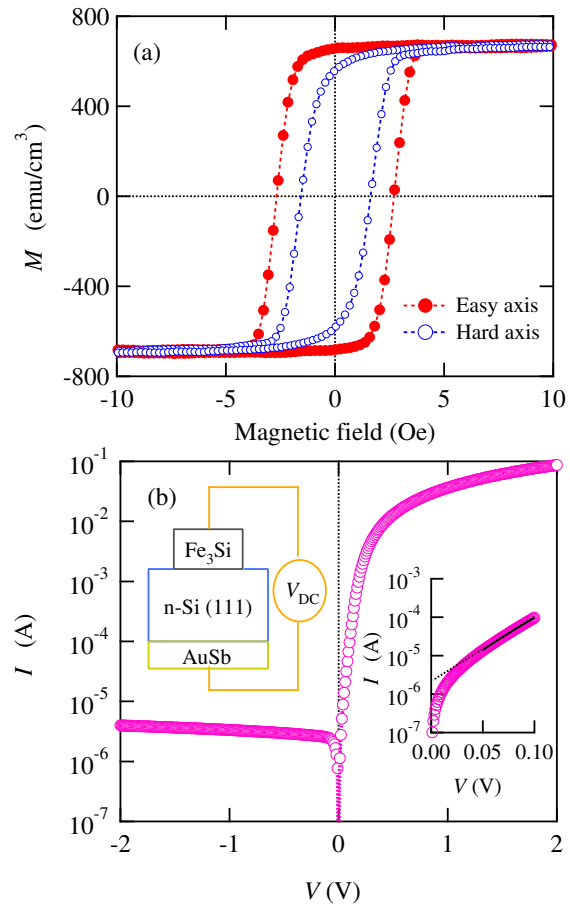}
\caption{(Color online) (a) Representative magnetization curves of the Fe$_{3}$Si/Si(111), measured along the magnetic easy and hard axes in the film plane at room temperature. (b) An $I-V$ characteristic of the Fe$_{3}$Si/Si(111) Schottky diode at room temperature. The insets show the schematic diagram of the fabricated Schottky diode and the $I-V$ characteristic in low forward bias regime.}
\end{figure}  

We also examine magnetic properties of the Fe$_{3}$Si/Si(111). Field-dependent magnetization measured at room temperature is shown in Fig. 3(a), where the applied field directions are parallel to the magnetic easy and hard axes in the film plane. Evident hysteretic and magnetic anisotropic features can be seen, showing ferromagnetic nature. We find that the saturation magnetization is estimated to be $\sim$ 650 emu/cm$^{3}$, i.e. $\sim$ 3.16 $\mu_\text{B}$/f.u. From the depth profile measurements of Fe and Si atoms by energy dispersive x-ray spectroscopy and RBS, we determined the chemical composition of our sample as Fe$_{2.84}$Si$_{1.16}$. On the other hand, Lenz {\it et al.} have already reported that the saturation magnetization is $\sim$ 790 emu/cm$^{3}$, i.e., $\sim$ 3.84 $\mu_\text{B}$/f.u. for an epitaxial Fe$_{3}$Si thin film with stoichiometric atomic composition.\cite{Lenz} From these facts, since the Fe content of our sample is slightly lower than that of the stoichiometric samples, the lack of the magnetization can be explained by an influence of a small amount of the disordered phases.\cite{Ionescu,Moss,Ploog,Nakane} 

Finally, to further evaluate the electrical properties of the Fe$_{3}$Si/Si(111) interface, we fabricated Fe$_{3}$Si/Si(111)/AuSb Schottky diodes, as shown in the left inset of Fig. 3(b), and measured the $I-V$ characteristics. Here, a resistivity of the used n-type Si(111) substrate is 1 $\sim$ 5 $\Omega$ cm (an impurity concentration of $\sim$ 10$^{15}$ cm$^{-3}$). We display absolute values of current as a function of bias voltage in the main panel of Fig. 3(b), where this behavior is reproduced for ten devices. A rectifying behavior indicating thermionic emission electron transport across a Schottky barrier can be observed, and the ratio of the current in forward bias to that in reverse bias is above $\sim$ 10$^{4}$. From fitting the data in the forward bias (0.05 V $\lesssim$ $V$ $\lesssim$ 0.1 V, see right inset),\cite{ref2} the Schottky barrier height ($\phi_\text{B}$) can roughly be estimated to be $\sim$ 0.62 eV, being in good agreement with $\phi_\text{B}$ obtained by its $C-V$ characteristic measurements. Note that the ideality factor $n$ becomes $\sim$ 1.08. These mean that the Fe$_{3}$Si/Si(111) interface also demonstrates ideal electrical properties without interfacial defects.

In summary, we have realized epitaxial growth of ferromagnetic Fe$_{3}$Si layers on Si (111) with keeping an abrupt interface using low-temperature molecular beam epitaxy at 130 $^{\circ}$C. The grown Fe$_{3}$Si layers have the ordered $DO_\text{3}$ phase. We believe that development of this work can open up a road to operations of silicon-based spintronic devices with a spin injector and detector using the Fe$_{3}$Si/Si interfaces. 
\vspace{3mm}

The authors thank Prof. Y. Maeda of Kyoto University and Prof. K. Matsuyama and Prof. Y. Nozaki of Kyushu University for useful discussion and providing the opportunity to use a VSM. K.U. and Y.A. acknowledge support from JSPS Research Program for Young Scientists. This work was partly supported by a Grant-in-Aid for Scientific Research on Priority Area (No.18063018) from the Ministry of Education, Culture, Sports, Science, and Technology in Japan.









\begin{thebibliography}{11}
\bibitem{Wolf}
S. A. Wolf, D. D. Awschalom, R. A. Buhrman, J. M. Daughton, S. v. Moln\'ar, M. L. Roukes, A. Y. Chtchelkanova, and D. M. Treger, Science {\bf 294}, 1488 (2001).
\bibitem{Datta}
S. Datta and B. Das, Appl. Phys. Lett. {\bf 56}, 665 (1990).
\bibitem{Sugahara}
S. Sugahara and M. Tanaka, Appl. Phys. Lett. {\bf 84}, 2307 (2004).
\bibitem{Hall}
K. C. Hall and M. E. Flatt\'e, Appl. Phys. Lett. {\bf 88}, 162503 (2006). 
\bibitem{Hanbicki}
A. T. Hanbicki, B. T. Jonker, G. Itskos, G. Kioseoglou, and A. Petrou, Appl. Phys. Lett {\bf 80}, 1240 (2002).
\bibitem{Hanbicki2}
A. T. Hanbicki, O. M. J. van't Erve, R. Magno, G. Kioseoglou, C. H. Li, B. T. Jonker, G. Itskos, R. Mallory, M. Yasar, and A. Petrou, Appl. Phys. Lett {\bf 82}, 4092 (2003).
\bibitem{Lou}
X. Lou, C. Adelmann, S. A. Crooker, E. S. Garlid, J. Zhang, S. M. Reddy, S. D. Flexner, C. J. Palmstr\o m, and P. A. Crowell, Nature Physics {\bf 3}, 197 (2007).
\bibitem{Min}
B. Min, K. Motohashi, C. Lodder, and R. Jansen, Nat. Mater. {\bf 5}, 817 (2006); R. Jansen, Nat. Phys. {\bf 3}, 521 (2007).
\bibitem{Appelbaum}
I. Appelbaum B. Huang, and D. J. Monsma, Nature {\bf 447}, 295 (2007).
\bibitem{Jonker1}
B. T. Jonker, G. Kioseoglou, A. T. Hanbicki, C. H. Li, and P. E. Thompson, Nature Physics {\bf 3}, 542 (2007).
\bibitem{Jonker2}
O. M. J. van't Erve, A. T. Hanbicki, M. Holub, C. H. Li, C. Awo-Affouda, P. E. Thompson, and B. T. Jonker, Appl. Phys. Lett. {\bf 91}, 212109 (2007).
\bibitem{Tanaka}
K. Akeura, M. Tanaka, M. Ueki, and T. Nishinaga, Appl. Phys. Lett. {\bf 67}, 3349 (2005).
\bibitem{Plake}
T. Plake, T. Hesjedal, J. Mohanty, M. K\"astner, L. D\"aweritz, and K. H. Ploog,  Appl. Phys. Lett. {\bf 82}, 2308 (2003).
\bibitem{Ionescu}
A. Ionescu, C. A. F. Vaz, T. Trypiniotis, C. M. G\"urtler, H. Garc\'{\i}a-Miquel, J. A. C. Bland, M. E. Vickers, R. M. Dalgliesh, S. Langridge, Y. Bugoslavsky, Y. Miyoshi, L. F. Cohen, and K. R. A. Ziebeck, Phys. Rev. B {\bf 71}, 094401 (2005).
\bibitem{ref1}
Although the Si(222) peak is forbidden reflection theoretically, the small peak can be observed for actual Si substrates. Simultaneously, a quite small peak resulted from the Fe$_{3}$Si(222) peak can also be detected near the Si(222) peak. 
\bibitem{Sadoh}
T. Sadoh, M. Kumano, R. Kizuka, K. Ueda, A. Kenjo, and M. Miyao, Appl. Phys. Lett. {\bf 89}, 182511 (2006); K. Ueda, Y. Ando, M. Kumano, T. Sadoh, Y. Maeda, and M. Miyao, Appl. Sur. Sci. {\bf 254}, 6215 (2008).
\bibitem{Maeda}
Y. Maeda, T. Jonishi, K. Narumi, Y. Ando, K. Ueda, M. Kumano, T. Sadoh, and M. Miyao, Appl. Phys. Lett. {\bf 91}, 171910 (2007). 
\bibitem{Lenz}
K. Lenz, E. Kosubek, K. Baberschke, H. Wende, J. Herfort, H. P. Sch\"onherr, and K. H. Ploog, Phys. Rev. B {\bf 72}, 144411 (2005).
\bibitem{Moss}
J. Moss and P. J. Brown, J. Phys. F: Metal Phys. {\bf 2}, 358 (1972).
\bibitem{Ploog}
J. Herfort, H. -P. Sch\"onherr, K.-J. Friedland, and K. H. Ploog, J. Vac. Sci. Technol. B {\bf 22}, 2073 (2004).
\bibitem{Nakane}
R. Nakane, M. Tanaka, and S. Sugahara, Appl. Phys. Lett. {\bf 89}, 192503 (2006). 
\bibitem{ref2}
Following the thermionic emission model, we used next relationships, $I =$ $I_\text{S}$($e$$^{qV/nkT}$$-$ 1), where $I_\text{S} =$ $A_{e}$$A^{*}$$T^{2}$$e$$^{-q\phi_\text{B}/kT}$. $A_{e}$ is the junction area of 7.6 $\times$ 10$^{-3}$ cm$^{2}$ and $A^{*}$ is the Richardson constant (110 A/cm$^{2}$K$^{2}$).



\end{thebibliography}
\end{document}